\documentclass[sigconf,screen,nonacm]{acmart}

\setcopyright{none}
\renewcommand\footnotetextcopyrightpermission[1]{}
\usepackage{booktabs}
\usepackage{amsmath}

\usepackage{amssymb}
\usepackage{algorithmic}
\usepackage{graphicx}
\usepackage{textcomp}
\usepackage[table,xcdraw]{xcolor}
\usepackage{fancybox}
\usepackage[most]{tcolorbox}
\usepackage{multirow}

\AtBeginDocument{%
}

\begin{document}

\title{A Pilot Study on Curator-Guided Multilingual Art Description for Blind and Low-Vision Audiences with Small Vision--Language Models}

\author{Iosif Tsangko}
\email{iosif.tsangko@tum.de}
\affiliation{%
  \institution{Technical University of Munich}
  \city{Munich}
  \country{Germany}
}

\author{Andreas Triantafyllopoulos}
\affiliation{%
\institution{Technical University of Munich}
\city{Munich}
\country{Germany}
}

\author{George Margetis}
\affiliation{%
\institution{Foundation for Research and Technology -- Hellas}
\city{Heraklion}
\country{Greece}
}

\author{Ioana Crihana}
\affiliation{%
\institution{National University of Science and Technology Politehnica Bucharest}
\city{Bucharest}
\country{Romania}
}

\author{Björn W. Schuller}
\affiliation{%
  \institution{Technical University of Munich}
  \city{Munich}
  \country{Germany}
}

\begin{abstract}
Artificial intelligence (AI) can vastly improve content personalisation for diverse users.
In the space of art, in particular, blind and low-vision (BLV) individuals are currently excluded from most visual content.
To mitigate this, museums need secure, on-premise vision-language models (VLMs) to produce high-quality audio descriptions for BLV audiences across languages, including low-resource ones.
In this work, we explore different design choices for small VLMs: under a fixed backbone and training budget, do language-specific adapters outperform a single multilingual adapter?
Using Qwen2.5-VL-3B-Instruct, we build a parallel BLV-oriented caption corpus in German, Romanian, and Serbian, and fine-tune monolingual and multilingual adapters. We evaluate both regimes with automatic metrics and an LLM-as-Judge protocol spanning multiple state-of-the-art models, and calibrate judge reliability with a Romanian BLV pilot study by measuring agreement between human and LLM preferences. Under our pilot setup, language-specific adapters show more stable controllability and visually grounded description quality for Romanian and Serbian, while multilingual adaptation remains competitive in German. We interpret these findings as deployment-oriented evidence for small on-premise VLMs.

\end{abstract}




\maketitle
%tables: https://www.overleaf.com/project/687e3967b77e91c7c7b6c247

\section{Introduction}

Foundation models are increasingly used as general-purpose multimedia backbones in
high-stakes settings~\cite{bommasani21-FM,nist23-AIRMF,tsangko25-RSP,triantafyllopoulos25-JPROC, triantafyllopoulos24-MH,schuller24-ACH}, which raises
practical questions about deploying capable models under strict privacy and infrastructure
constraints. In museums and art institutions, the mandate to protect visitor privacy often necessitates \emph{on-premise} inference, while the desire for controllability incentivises \emph{open-weight}
models~\cite{eu24-AIACT}.
These requirements raise the need for compact vision-language models (VLMs) and parameter-efficient adaptation.

Prior BLV-oriented captioning work has largely focused on assistive, in-the-wild scenarios
(e.g., photos taken by blind users), where the main objective is immediate utility via
object naming and short factual statements. In contrast, museum descriptions require
longer, \emph{orientation-first} narratives with stable structure and controllable verbosity,
often closer to curated description practice than to generic image captioning
\cite{Snyder14-VMV,Kleege15-ADA,Salisbury17-TSS,Perego19-ITL,Gurari20-CIT, Alharbi24-MWA}.

From a modelling perspective, lightweight adapters such as LoRA~\cite{hu21-LRA} make small VLMs practical
without full fine-tuning, while multilingual multimodal resources
(e.g., Multi30K~\cite{Elliott16-MME} and WIT~\cite{Srinivasan21-WWI}) have enabled cross-lingual training beyond English. However, for multilingual BLV museum description at
the $\sim$3B scale, the deployment choice between sharing a \emph{single multilingual}
adapter versus allocating the same capacity to \emph{language-specific} adapters remains
underexplored, especially for lower-resource 
languages~\cite{Chu23-MAF,Chu24-MVV,Rust21-HGI,Clark22-CPA,Goyal22-TFE}.

Finally, caption evaluation is increasingly recognised as under-specified by $n$-gram metrics alone~\cite{callisonburch06-BLEU,papineni02-BLEU,novikova17-NEW,lin04-ROUGE}.
Rubric-driven, model-based evaluation (LLM-as-Judge) has gained traction as a scalable alternative~\cite{zheng23-MTB,chiang24-ARENA,dubois24-ALP2,wanga23-CHATGPT-EVAL,kocmi23-MTEVAL},
but must first be validated against human ratings to draw robust conclusions~\cite{liu23-GEV,reiter09-INV}. Motivated by this, we complement standard automatic metrics with a BLV
rubric and calibrate candidate judges on a small BLV pilot before scaling evaluation to
languages where BLV annotations are unavailable.

Beyond a language-only comparison, our contribution is inherently multimedia: we study accessible description generation from the combination of \emph{artwork images}, \emph{curatorial metadata}, and \emph{language-specific descriptive conventions} in a realistic museum deployment setting. The paper therefore addresses a multimodal problem at three levels: multimodal input (image + metadata), multimodal generation for accessibility (visual content rendered into structured BLV-oriented language), and multimodal evaluation, where lexical metrics are complemented by visually grounded rubric-based judging. Under this framing, the comparison is not an isolated NLP question, but a deployment design study for compact on-premise multimedia systems that must remain controllable, privacy preserving, and usable in low-resource languages.

\begin{figure*}[h]
  \centering
  \includegraphics[width=0.9\textwidth]{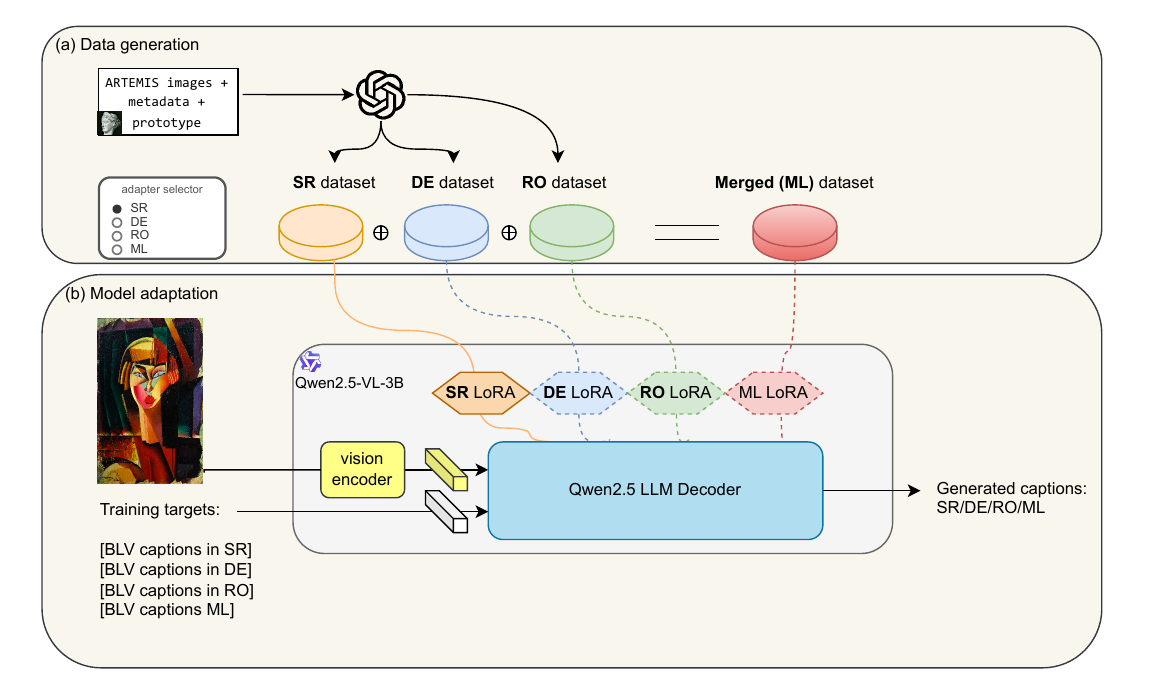}
  \caption{End-to-end pipeline. ARTEMIS artwork images and metadata (style, author/work IDs, emotion) feed curator-written BLV prototypes (structure–verbosity–orientation). A vision LLM generates one-shot, language-specific pools (DE/SR/RO), followed by post-processing (length cap, orientation cues, object order, landmarks). These corpora fine-tune small VLMs (monolingual and multilingual).}
  \label{fig:blv-pipeline}
\end{figure*}

\subsection{Research Questions}
We study \emph{small, open-weight, on-premise} VLMs (approximately 3B parameters)
for BLV art description in German (DE), Romanian (RO), and Serbian (SR),
with a held-out test split consisting of \emph{artwork styles} not seen in training. Our work investigates the following research questions:

\textbf{RQ1 (Adapter strategy under a fixed budget).}
Given the same backbone and LoRA parameter budget, does \emph{language-specific} adaptation (one adapter per language)
outperform a \emph{single multilingual} adapter for BLV description quality and controllability, and how do both regimes
compare to the base model (zero-shot)?

\vspace{2pt}
\textbf{RQ2 (Robustness across languages).}
Does multilingual parameter sharing change robustness and controllability relative to language-specific adapters across DE, RO, and SR on unseen styles, and where do the regimes differ most in error type (length, repetition, and visually grounded content)?

\vspace{2pt}
\textbf{RQ3 (Human-grounded evaluation at scale).}
After calibrating an LLM-as-Judge rubric on a Romanian BLV pilot study, can LLM judges reliably reproduce human
\emph{mono-multi} preferences and trait-level score differences (notably for visually grounded vs affective traits),
so that evaluation can be extended to DE and SR where BLV annotations are unavailable?

\begin{table*}[ht]
\centering
\small
\caption{Overview of the English base dataset and derived BLV caption corpora.
Top blocks: ART-GenEvalGPT subset, pruning, and style-held-out split with
per-emotion counts. Bottom block: per-language caption corpora and basic text
statistics (train means).  Here, \emph{desc} denotes one BLV
description; \emph{Words/desc}, \emph{Sent./desc}, and \emph{Words/sent.}
are the mean number of words per description, sentences per description, and
words per sentence, respectively (reported as Tr/Te).}
\label{tab:data-overview}
\begin{tabular}{lccccccc}
\toprule
\textbf{Category / Corpus} &
\textbf{\# Items} &
\textbf{Train} &
\textbf{Test} &
\textbf{Notes} &
\textbf{Words/desc} &
\textbf{Sent./desc} &
\textbf{Words/sent.} \\
\midrule
\multicolumn{8}{l}{\textit{Dataset stages and filtering (English base)}} \\
\cmidrule(lr){1-8}
Safe dialogues (initial)      & 799 & --  & --  & English; image, metadata          & --     & --    & --    \\
After pruning                 & 765 & --  & --  & Drop rare styles $<\!10$ & --     & --    & --    \\
Train split                   & 553 & 553 & --  & All styles except held-out                 & --     & --    & --    \\
Test split (style-held-out)        & 212 & --  & 212 & Impres., Expres., Na\"{i}ve/Primitivism & -- & --    & --    \\
\midrule
\multicolumn{8}{l}{\textit{Per-emotion counts after pruning and style-held-out split}} \\
\cmidrule(lr){1-8}
amusement                     & 81  &  58 & 23  & -- & -- & -- & -- \\
anger                         & 66  &  55 & 11  & -- & -- & -- & -- \\
awe                           & 92  &  86 &  6  & -- & -- & -- & -- \\
contentment                   & 84  &  42 & 42  & -- & -- & -- & -- \\
disgust                       & 61  &  42 & 19  & -- & -- & -- & -- \\
excitement                    & 55  &  39 & 16  & -- & -- & -- & -- \\
fear                          & 86  &  64 & 22  & -- & -- & -- & -- \\
neutral                       & 154  & 111 & 43  & -- & -- & -- & -- \\
sadness                       & 86  &  56 & 30  & -- & -- & -- & -- \\
\cmidrule(lr){1-8}
\textbf{Totals}               & 765  & \textbf{553} & \textbf{212} & -- & -- & -- & -- \\
\midrule
\multicolumn{8}{l}{\textit{BLV caption corpora (generated in DE, SR, RO)}} \\
\cmidrule(lr){1-8}
German (DE)                   & 765  & 553 & 212 & Caption corpus & 263.39 & 15.36 & 17.29 \\
Serbian (SR)                  & 765  & 553 & 212 & Caption corpus & 215.45 & 12.43 & 17.52 \\
Romanian (RO)                 & 765  & 553 & 212 & Caption corpus & 224.85 & 11.05 & 20.54 \\
\textbf{Multilingual (DE+SR+RO)} &
\textbf{2295} & \textbf{1659} & \textbf{636} & Merged DE+SR+RO &
\textbf{$\sim$234.6} & \textbf{$\sim$12.95} & \textbf{$\sim$18.45} \\
\bottomrule
\end{tabular}
\end{table*}

\subsection{Experiment Scope and Model Choices}
To isolate the mono vs multilingual effect under realistic museum constraints,
we fix a single small open-weight VLM backbone: Qwen2.5-VL-3B-Instruct\footnote{https://huggingface.co/Qwen/Qwen2.5-VL-3B-Instruct}. 
We chose this model because it is (i) small enough for on-premise deployment,
(ii) trained for multilingual captioning with a strong vision encoder, and
(iii) fully open, which makes replication and downstream use straightforward.
All language-specific and multilingual adapters are implemented with LoRA layers added to this backbone.

For evaluation beyond automatic lexical and embedding metrics, we employ an
LLM-as-Judge setup. We first calibrate several candidate judges (e.g., GPT-4o,
Claude, Gemma-3-12B-IT\footnote{https://huggingface.co/google/gemma-3-12b-it})
against a small set of Romanian BLV human annotations and then use the best-aligned judges
to compare monolingual and multilingual systems across all three languages.
Details of this calibration are given in the results section.

\section{Methodology}

\subsection{Dataset}
We build on ARTEMIS~\cite{achlioptas21-AAL}, an image–emotion corpus linking artworks to human affect annotations, and its curated, safety-screened derivative Art-GenEvalGPT \cite{gil24-ADO}. From the latter, we retain only entries flagged as safe dialogues, together with minimal metadata required for our task: artwork/style codes, author and artwork identifiers, emotion labels, and image paths. This yields 799 image–dialogue pairs (English) that serve as the source content pool from which we later derive language-specific accessible descriptions. 
To obtain a balanced, well-specified base for training/evaluation prior to multilingual generation, we apply two lightweight steps:

\textit{Pruning rare categories}. We drop styles with fewer than 10 samples and remove the underspecified emotion label ``something else''. This reduces the set to 765 items.

\textit{Style-held-out split.} To test generalisation across visual domains, we construct a style-held-out
%BS: added "held" after style also in another place...
evaluation: three highly entangled styles -- Impressionism (often contentment-leaning), Expressionism (fear-leaning), and Naive Art/Primitivism (amusement-leaning) -- are reserved for test. The remaining styles form train. This yields 553 train and 212 test samples, maintaining diverse emotion distributions in both splits as in Table~\ref{tab:data-overview}. 

\emph{Data Statistics}. Each English base item yields one accessibility-oriented description per target language. We achieve this by producing one description per language using GPT-4o-mini\footnote{https://platform.openai.com/docs/models/gpt-4o-mini, snapshot: gpt-4o-mini-2024-07-18} with the corresponding description with the artwork's metadata.
This produces three monolingual corpora whose sizes mirror the base split (553 train / 212 test per language). A multilingual corpus is formed by concatenating the three monolingual sets and shuffling (1,659 train / 636 test). 
We report simple, language-agnostic text metrics (means per description): words, sentences, words/sentence, and type–token ratio (TTR). 
For details see Table~\ref{tab:data-overview}.

A key design choice in our pipeline is that supervision is \emph{prototype-guided} rather than prompt-only. For each target language, curator-provided BLV prototype descriptions define the desired structure, level of detail, and orientation-first style, and are combined with shared accessibility instructions, artwork metadata, and the artwork image to generate one synthetic reference description per item. These references are therefore intended to operationalise a \emph{target descriptive format} grounded in professional practice, not to serve as universal ground truth for art interpretation. Accordingly, training and reference-based evaluation should be read primarily as measuring how well a model reproduces this target BLV description style; that is, controllability and distillation fidelity under a fixed multimedia generation protocol.

\subsection{Training}
We train the VLM in four regimes: three strictly monolingual runs (SR, DE, RO) and one multilingual run trained on the concatenation of the three languages. The multilingual run shuffles the merged pool. In all cases, the model is trained autoregressively on instruction–response samples that pair the artwork image with a BLV-oriented target description generated as described, preserving a consistent schema across languages.

\begin{table*}[t]
\centering\small
\caption{Comprehensive results across metrics. Embedding similarity is cosine with \texttt{gte-multilingual-base} (higher is better).
Length error is mean absolute character difference (lower is better). Repetition is: $1-$token-level ratio (lower is better).
Lexical metrics are ROUGE-L, BLEU, and chrF (higher is better).
LS = language-specific (\textbf{mono}lingual); ML = \textbf{multi}lingual.
$\Delta$ denotes $\Delta_{\text{LS-ML}} = \text{LS} - \text{ML}$ (positive favours LS).
\textbf{The base model uses the same backbone in both columns; hence LS=ML and $\Delta=0$ by definition.}
}
\label{tab:all-metrics}
\setlength{\tabcolsep}{4pt}
\resizebox{\textwidth}{!}{
\begin{tabular}{llrrrrrrrrrrrrr}
\toprule
& & \multicolumn{3}{c}{\textbf{Embedding cosine}~$\uparrow$} & \multicolumn{2}{c}{\textbf{Length error}~$\downarrow$} & \multicolumn{2}{c}{\textbf{Repetition}~$\downarrow$} & \multicolumn{2}{c}{\textbf{ROUGE-L}~$\uparrow$} & \multicolumn{2}{c}{\textbf{BLEU}~$\uparrow$} & \multicolumn{2}{c}{\textbf{chrF}~$\uparrow$} \\
\cmidrule(lr){3-5}\cmidrule(lr){6-7}\cmidrule(lr){8-9}\cmidrule(lr){10-11}\cmidrule(lr){12-13}\cmidrule(lr){14-15}
\textbf{Lang.} & \textbf{Variant}
& {LS} & {ML} & {$\Delta$}
& {LS} & {ML}
& {LS} & {ML}
& {LS} & {ML}
& {LS} & {ML}
& {LS} & {ML} \\
\midrule
\multirow{2}{*}{German}
  & Base model               & 0.686 & 0.686 & +0.000 & 432 & 432 & 0.378 & 0.378 & 0.162 & 0.162 & 0.041 & 0.041 & 0.380 & 0.380 \\
  & \textbf{Fine-tuned}     & \textbf{0.799} & \textbf{0.688} & \textbf{+0.111} & 229 & 276 & 0.376 & \textbf{0.345} & 0.160 & \textbf{0.209} & 0.090 & \textbf{0.090} & 0.405 & \textbf{0.449} \\[2pt]
\multirow{2}{*}{Romanian}
  & Base model               & 0.651 & 0.651 & +0.000 & 386 & 386 & 0.386 & 0.386 & 0.146 & 0.146 & 0.028 & 0.028 & 0.333 & 0.333 \\
  & \textbf{Fine-tuned}     & \textbf{0.796} & \textbf{0.696} & \textbf{+0.100} & 209 & 541 & \textbf{0.323} & \textbf{0.261} & \textbf{0.228} & \textbf{0.225} & \textbf{0.117} & 0.075 & \textbf{0.466} & \textbf{0.353} \\[2pt]
\multirow{2}{*}{Serbian}
  & Base model               & 0.634 & 0.634 & +0.000 & 516 & 516 & 0.509 & 0.509 & 0.001 & 0.001 & 0.014 & 0.014 & 0.262 & 0.262 \\
  & \textbf{Fine-tuned}     & \textbf{0.814} & \textbf{0.746} & \textbf{+0.068} & 220 & \textbf{207} & \textbf{0.311} & \textbf{0.295} & \textbf{0.564} & 0.101 & \textbf{0.097} & 0.095 & \textbf{0.425} & 0.419 \\
\bottomrule
\end{tabular}}
\end{table*}

Optimisation follows a lightweight LoRA setup over all linear modules with $16-bit$ floating point activations: 3 epochs, per-device batch size 2 for train and eval, learning rate $10^{-4}$, context length $2048$ tokens, and periodic evaluation every 25 steps with checkpointing every 100 steps. Adapters use rank 132 and $\alpha=32$. Representative runs expose $\approx$247M trainable parameters (about 6\% of the full backbone, as reported by our training setup). Best checkpoints are selected by held-out eval loss/accuracy (monolingual: step 100–105; multilingual: step 519). 

% We keep preprocessing and prompting identical across regimes and do not mix languages within a sample in the multilingual run. This isolates the effect of parameter sharing across languages from confounds due to format drift. Any metadata-augmented variants (artist/style, emotion) follow the same training recipe and differ only in the presence of those fields in the input prompt. 

% \begin{figure*}[t]
%   \centering
%   % Prefer a vector PDF; replace filename if needed.
%   \includegraphics[width=\textwidth]{figures/monvsmulti.pdf}
%   \caption{t-SNE maps of document embeddings (GTE-Multilingual-Base) for German (top), Romanian (middle), and Serbian (bottom).
%   Left column: \emph{language-specific} adapters; right column: \emph{multilingual} adapter.
%   Points are coloured by variant (Reference, Zero-shot, Fine-tuned+metadata, Fine-tuned); translucent ellipses show 1\,SD contours and ✕ markers denote cluster centroids.
%   Panel titles include the silhouette coefficient; higher overlap with the reference ellipsoid indicates better alignment.}
%   \label{fig:tsne}
% \end{figure*}

\subsection{Experiments and metrics}
We compare two inference variants on the held-out split for each language:
\emph{Base model (zero-shot; no LoRA)} (base Qwen model with image+prompt) and \emph{Fine-tuned} (LoRA-adapted, image+prompt).
For every variant we evaluate both \emph{Language-Specific} (LS; monolingual) and
\emph{Multilingual} (ML; merged/multilingual) checkpoints. Metrics are computed per image
and averaged across the test set.

% Serbian outputs are NFC-normalised and script-checked (Cyrillic) before scoring.

\emph{Embedding similarity and projections.}
We embed references and system outputs with the multilingual embedding model \textit{gte multilingual base}\footnote{https://huggingface.co/Alibaba-NLP/gte-multilingual-base} (which supports text lengths up to 8192 tokens and over 70 languages) and report cosine similarity as a language-agnostic semantic score. 
% For qualitative structure we project the same embeddings to 2-D with t-SNE (perplexity 30, 1{,}000 iters, fixed seed) and visualise four clouds—Reference, Zero-shot, Fine-tuned, Fine-tuned+RAG—with centroids and 95\% covariance ellipses; titles include the mean silhouette coefficient over these four labels. These plots illustrate cluster alignment toward the reference space and the effect of retrieval on dispersion.

\emph{Lexical fidelity.}
We report BLEU-4, ROUGE-L, and chrF++ to capture surface overlap and ordering. chrF++ is included to reduce sensitivity to inflectional variation across the languages. Tokenisation and casing are held constant across LS/ML; we do not apply length penalties beyond the metric defaults.

% \emph{Semantic similarity (context-robust).}
% Beyond the GTE cosine, we compute BERTScore with \textit{xlm-roberta-large}\footnote{https://huggingface.co/FacebookAI/xlm-roberta-large} (multilingual) with baseline rescaling, reporting P/R/F1. 
% % We additionally compute MoverScore to probe paraphrase tolerance. 
% These metrics better correlate with curator judgements of BLV suitability than purely lexical measures.

\emph{Structure and format control.}
To track accessibility-oriented formatting, we measure (i) absolute character-length error
between candidate and reference (lower is better) and (ii) token repetition
$1-\frac{|\text{unique tokens}|}{|\text{tokens}|}$ (lower is better), which penalises
looping and redundant phrasing.

\emph{LLM-as-Judge (BLV rubric).}
We calibrate three candidate LLM judges, i.e., Gemma-3-12B-IT\footnote{\url{https://huggingface.co/google/gemma-3-12b-it}}, 
GPT-4o\footnote{\url{https://platform.openai.com/docs/models/gpt-4o}}, and
Claude~3.7~Sonnet\footnote{\url{https://www.anthropic.com/news/claude-3-7-sonnet}}, against the Romanian pilot study
(2 BLV raters, 25 items). Trait scores are produced with DeepEval's \texttt{GEval}
module\footnote{\url{https://deepeval.com/}}, an implementation of the rubric-driven
G-Eval protocol for LLM-based evaluation\cite{liu23-GEV}.
For each artwork and trait, we compute the per-item LS-ML difference
$\Delta = s^{\text{LS}} - s^{\text{ML}}$ and quantify judge-human alignment using
(i) Spearman rank correlation on $\Delta$ and (ii) \emph{sign agreement}, i.e., the
fraction of items where the judge and humans select the same winner (LS vs ML)
(Fig.~\ref{fig:llm-human-agreement}).

\emph{Pilot human evaluation (Romanian).}
We ran a small Romanian pilot to calibrate our metrics and LLM-as-Judge setup. Two native Romanian BLV participants evaluated 25 random test artworks in a custom Gradio interface, comparing two unlabelled captions (monolingual vs multilingual) with randomised left/right order. For each item, they rated both captions on five BLV traits (coherence, composition, colour fidelity, vividness, emotional tone; 1--10) and selected a preferred caption. This yields 500 trait ratings and 50 preference votes, collected independently.

\emph{Reporting.}
Unless otherwise stated, comparisons between LS and ML use the \emph{Fine-tuned} variant as the main line. 
% Models, prompts, and checkpoints will be released publicly upon acceptance.

% We release all trained adapters and checkpoints via our Hugging Face collection.\footnote{https://huggingface.co/collections/JoseferEins/shift-vision-language-models-for-accessible-ch}

\section{Results}

\paragraph{Embedding similarity.}
Fine-tuning substantially improves over the base model across all languages
(Table~\ref{tab:all-metrics}). Gains are largest for Romanian and Serbian,
where language-specific adapters show the strongest improvements, while German
exhibits smaller LS-ML gaps.
\paragraph{Surface form and control metrics.}Across all languages, LoRA fine-tuning improves both lexical overlap (ROUGE-L, BLEU, chrF) and output control (length error, repetition) over the base model (Table~\ref{tab:all-metrics}). Length error drops substantially for the LS adapters in German (432 to 229 chars) and Romanian (386 to 209 chars), with lower repetition in Romanian (0.386 to 0.323) and Serbian (0.509 to 0.311). Lexical gains are strongest in the lower-resource languages: Romanian LS improves from ROUGE-L 0.146 and BLEU 0.028 to 0.228 and 0.117 (chrF 0.466), and Serbian LS from ROUGE-L 0.001 and BLEU 0.014 to 0.564 and 0.097 (chrF 0.425). In German, LS and ML are closer on lexical metrics, with ML higher on ROUGE-L/chrF (0.209/0.449 vs 0.160/0.405) and equal BLEU (0.090), suggesting near-parity in this higher-resource setting.

\begin{figure*}[t]
\centering
\includegraphics[width=\textwidth]{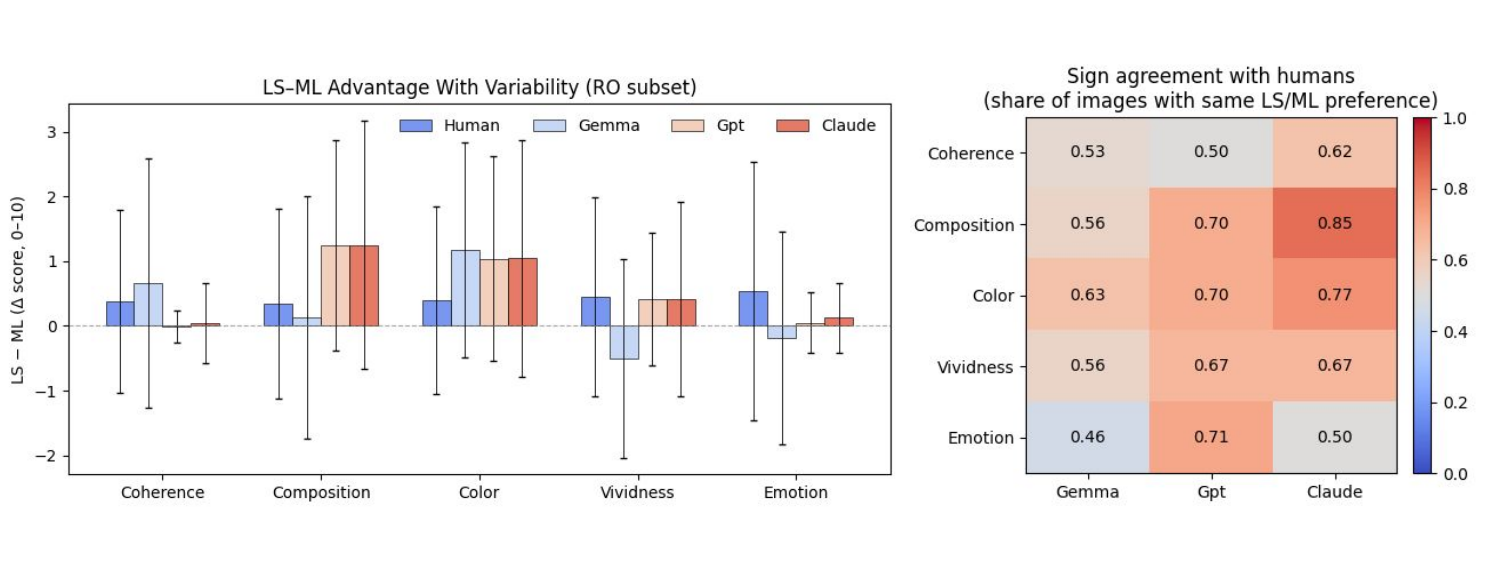}
\caption{
\textbf{LLM-as-Judge calibration on the Romanian (RO) pilot.}
\textbf{Left:} Mean LS-ML trait-score difference ($\Delta = s^{\text{LS}} - s^{\text{ML}}$; 0-10) with standard deviation across items, shown for humans and three LLM judges.
Positive values favour the language-specific (LS) adapter.
\textbf{Right:} Sign agreement with humans, i.e., the fraction of artworks where the judge and humans prefer the same system (LS vs ML) for a given trait.
Agreement is highest for visually grounded traits (composition, colour), especially for Claude and GPT-4o, while more subjective traits (vividness, emotion) show lower and more variable agreement; the smallest judge (Gemma) is least stable overall.
}
\label{fig:llm-human-agreement}
\end{figure*}

\paragraph{LLM-as-Judge calibration.}
Human inter-annotator agreement in the Romanian BLV pilot is moderate (Spearman
$\rho=0.32$-$0.49$ across traits), which is expected for subjective description-quality
ratings. To calibrate LLM judges, we compare monolingual
against multilingual captions by computing a per-item delta for each trait,
$\Delta_i = s_i^{\text{LS}} - s_i^{\text{ML}}$. We then quantify judge-human
alignment in two complementary ways: (i) Spearman rank correlation between human and
judge per-item deltas ($\Delta_i$), and (ii) \emph{sign agreement}, i.e., the fraction of
items where the judge and humans select the same winner (LS vs ML; Fig.~\ref{fig:llm-human-agreement}).

Alignment is strongest for visually grounded traits. For \emph{composition} and
\emph{colour fidelity}, Claude and GPT-4o best track the human per-item LS-ML pattern
($\rho=0.51$-$0.59$) and reach up to $85\%$ sign agreement, indicating that they often
identify the same images where LS improves (or degrades) factual visual content.
Agreement is weaker and less stable for more interpretive traits (\emph{vividness} and
\emph{emotional tone}), where both human deltas and judge deltas show higher dispersion,
suggesting that these dimensions admit multiple valid rubric interpretations at the
item level. Gemma shows the largest deviations from human preferences; this is plausible
given its 
%BS: unclear what "its" related to
smaller capacity, and we therefore treat it as a robustness check rather than
a basis for the main conclusions.

Overall, the pilot supports a trait-dependent view of judge validity: Claude is used as
the primary judge and GPT-4o as a secondary judge for DE/SR analyses, with Gemma retained
for additional diversity.

%BS: Fig. 2 - Color --> Colour (twice!), Gpt --> GPT

\paragraph{German and Serbian.}
We next apply the two calibrated judges (Claude and GPT-4o) to German (DE) and Serbian
(SR), where we do not have BLV human ratings. Table~\ref{tab:de_sr_mono_multi} reports
the mean LS-ML score difference per trait, $\Delta = s^{\text{LS}}-s^{\text{ML}}$,
together with $p_{\text{LS}}$ (the fraction of images where LS scores higher).

In German, both judges attribute a clear LS advantage to the most visually grounded
dimensions. For \emph{composition}, LS improves by $+1.58$ (Claude) and $+2.75$ (GPT-4o),
with LS winning on most items ($p_{\text{LS}}=0.75$ and $0.96$). For \emph{colour fidelity},
the gains are similarly consistent ($+1.88$ and $+1.96$), with LS preferred on at least
two thirds of the images (Claude: $0.67$; GPT-4o: $1.00$). In contrast, \emph{coherence},
\emph{vividness}, and \emph{emotional tone} exhibit small deltas close to zero and mixed
win rates, indicating approximate parity between LS and ML on these more interpretive
traits in DE.

Serbian shows the same main pattern for \emph{composition}: LS is preferred on average
($\Delta=+0.96$ for Claude; $+2.34$ for GPT-4o), and LS wins on a large share of items
($p_{\text{LS}}=0.67$ and $1.00$). For \emph{colour fidelity}, both judges again favour LS
but with stronger judge dependence (Claude: $+0.71$, $p_{\text{LS}}=0.54$; GPT-4o: $+2.07$,
$p_{\text{LS}}=0.92$). By contrast, \emph{coherence}, \emph{vividness}, and \emph{emotional
tone} have negative mean deltas in both judges, suggesting a mild ML tendency on these
dimensions; however, the effect sizes are small (all $|\Delta|<0.5$), so we interpret
them as secondary relative to the robust LS gains on factual traits.

Taken together, the DE/SR results mirror the Romanian calibration: LS adaptation most
reliably improves visually grounded content (composition, colour), while ML is competitive
on higher-level, stylistic dimensions where preferences are weaker and more variable.

\begin{table}[t]
\centering
\small
\begin{tabular}{llrrrr}
\toprule
 & & \multicolumn{2}{c}{Claude} & \multicolumn{2}{c}{GPT-4o} \\
\cmidrule(lr){3-4} \cmidrule(lr){5-6}
Language & Trait
         & $\Delta$ & $p_{\text{LS}}$
         & $\Delta$ & $p_{\text{LS}}$ \\
\midrule
\multirow{5}{*}{DE} 
& Coherence   & -0.04 & 0.04 & -0.09 & 0.46 \\
& Composition & \textbf{1.58} & \textbf{0.75} & \textbf{2.75} & \textbf{0.96} \\
& Colour       & \textbf{1.88} & \textbf{0.67} & \textbf{1.96} & \textbf{1.00} \\
& Vividness   & 0.29  & 0.33 & -0.05 & 0.54 \\
& Emotion     & -0.04 & 0.08 & -0.18 & 0.38 \\
\midrule
\multirow{5}{*}{SR}
& Coherence   & \textbf{-0.25} & \textbf{0.17} & \textbf{-0.25} & \textbf{0.25} \\
& Composition & \textbf{0.96}  & \textbf{0.67} & \textbf{2.34}  & \textbf{1.00} \\
& Colour       & 0.71  & 0.54 & \textbf{2.07}  & \textbf{0.92} \\
& Vividness   & \textbf{-0.21} & \textbf{0.08} & \textbf{-0.43} & \textbf{0.29} \\
& Emotion     & \textbf{-0.21} & \textbf{0.12} & \textbf{-0.35} & \textbf{0.25} \\
\bottomrule
\end{tabular}
\caption{
LS-ML deltas for German (DE) and Serbian (SR) from calibrated LLM judges (Claude, GPT-4o).
$\Delta = s^{\text{LS}}-s^{\text{ML}}$ is the mean trait-score difference (positive favours LS).
$p_{\text{LS}}$ is the fraction of images where LS scores higher.
\textbf{Bold} marks consistent preferences ($|\Delta|>0.5$ and $p_{\text{LS}}\ge0.67$ or $\le0.33$).
}
\label{tab:de_sr_mono_multi}
\end{table}

\section{Discussion and Conclusion}
We addressed a practical museum deployment question for \emph{small, open-weight, on-premise} VLMs:
given a fixed backbone and LoRA budget, should BLV-oriented art description use a \emph{single multilingual}
adapter or \emph{separate monolingual} adapters?

\paragraph{Main findings.}
Across German, Romanian, and Serbian, LoRA fine-tuning consistently improves over the base model in both semantic
and surface-form metrics and reduces repetition. Under matched capacity in our pilot setup, \emph{language-specific} adapters appear more stable in the lower-resource settings, where they yield tighter controllability (verbosity and repetition) and stronger visually grounded quality.
In German, the multilingual adapter remains competitive, indicating that cross-lingual transfer can be sufficient
when the language is well covered by the base model.

\paragraph{Evaluation implications.}
Calibrating LLM-as-Judge on a Romanian BLV pilot enables scalable, trait-level comparison when BLV annotations
are unavailable. In our setup, stronger proprietary judges align best with BLV preferences for \emph{factual,
image-grounded} traits (e.g., composition and colour fidelity), while \emph{affective/stylistic} traits show lower
human agreement and weaker judge alignment; these dimensions should therefore be interpreted cautiously in
cross-language conclusions.
The Romanian BLV pilot should be interpreted as a \emph{calibration study}, not as a comprehensive user evaluation. Its role is to test whether the proposed rubric and candidate LLM judges capture the same \emph{relative} LS--ML preferences as human raters, especially for visually grounded traits such as composition and colour fidelity. It does \emph{not} establish broad cross-language accessibility effectiveness, nor does it replace larger BLV studies in German and Serbian. We therefore use the pilot to justify judge selection and to support cautious trait-level interpretation, while reserving stronger claims about end-user accessibility benefit for future, larger-scale human evaluation.
\paragraph{Limitations and next steps.}
Our BLV references are synthetically generated, so reference-based metrics can inherit teacher-student biases,
and automatic scores remain imperfect proxies for accessibility. The human study is a small Romanian pilot used
primarily for judge calibration; broader BLV evaluations in DE and SR are needed to confirm trait-level trends
and quantify practical effect sizes. We plan to release curated splits, prompts, and LS/ML adapters for
reproducibility.

\section*{Acknowledgements}

This work has received funding from the European Union's Horizon Europe research and innovation programme under grant agreement No. 101060660 (SHIFT).

\bibliographystyle{ACM-Reference-Format}
\bibliography{refs}

\end{document}